\documentclass[prb,preprint]{revtex4}
\usepackage{graphicx}
\usepackage{latexsym}
\usepackage{amsmath}
\usepackage{bm}
\newcommand{\beq}{\begin{equation}}
\newcommand{\eeq}{\end{equation}}
\newcommand{\ba}{\begin{array}{ccc}}
\newcommand{\ea}{\end{array}}
\newcommand{\nn}{\nonumber \\}

\newcommand{\br}{{\bm r}}
\newcommand{\bk}{{\bm k}}

\newcommand{\bq}{{\bm q}}

\newcommand{\bQ}{{\bm Q}}
\def\bea{\begin{eqnarray}}
\def\eea{\end{eqnarray}}
\begin{document}
\title{Auxiliary-boson and DMFT studies of bond ordering instabilities of $t$-$J$-$V$ models on the square lattice}

\author{Andrea Allais}
\affiliation{Department of Physics, Harvard University, Cambridge MA 02138}

\author{Johannes Bauer}
\affiliation{Department of Physics, Harvard University, Cambridge MA 02138}

\author{Subir Sachdev}
\affiliation{Department of Physics, Harvard University, Cambridge MA
02138}

\date{\today}

\begin{abstract}
We examine the influence of strong on-site Coulomb interactions on instabilities of the metallic state
on the square lattice to general forms of bond order. The Mott correlations are accounted
for by the auxiliary-boson method, and by dynamical mean field theory calculations, complementing our recent work 
(arXiv:1402.4807) using Gutzwiller projected
variational wavefunctions. By the present methods, 
we find that the on-site Mott correlations do not significantly modify the structure of the bond ordering
instabilities which preserve time-reversal symmetry, but they do enhance the instability
towards time-reversal symmetry breaking ``staggered flux'' states. 
\end{abstract}

\maketitle

\section{Introduction}

In a recent paper \cite{paper1}, we examined instabilities of $t$-$J$-$V$ models on the square lattice
to arbitrary orderings in the spin-singlet, particle-hole channel, and accounted for the on-site Coulomb interactions by
a variational wavefunction which projected out sites with double occupancy. 
In the present paper we will examine essentially the same models,
but will account for the on-site interactions by the auxiliary-boson method (also called the ``slave-boson'' method)
and dynamical
mean field theory (DMFT) calculations.
As in the previous work \cite{paper1}, our analysis allows for
 charged stripes,\cite{stripes} checkerboard and
bond density waves,\cite{ssrmp,vojta4,dhlee} 
Ising-nematic order,\cite{YK00,HM00,OKF01} staggered flux states,\cite{marston,kotliar,sudip,leewen,laughlin} and states with spontaneous currents.\cite{varma}

\begin{figure}[h]
\includegraphics[width=3.5in]{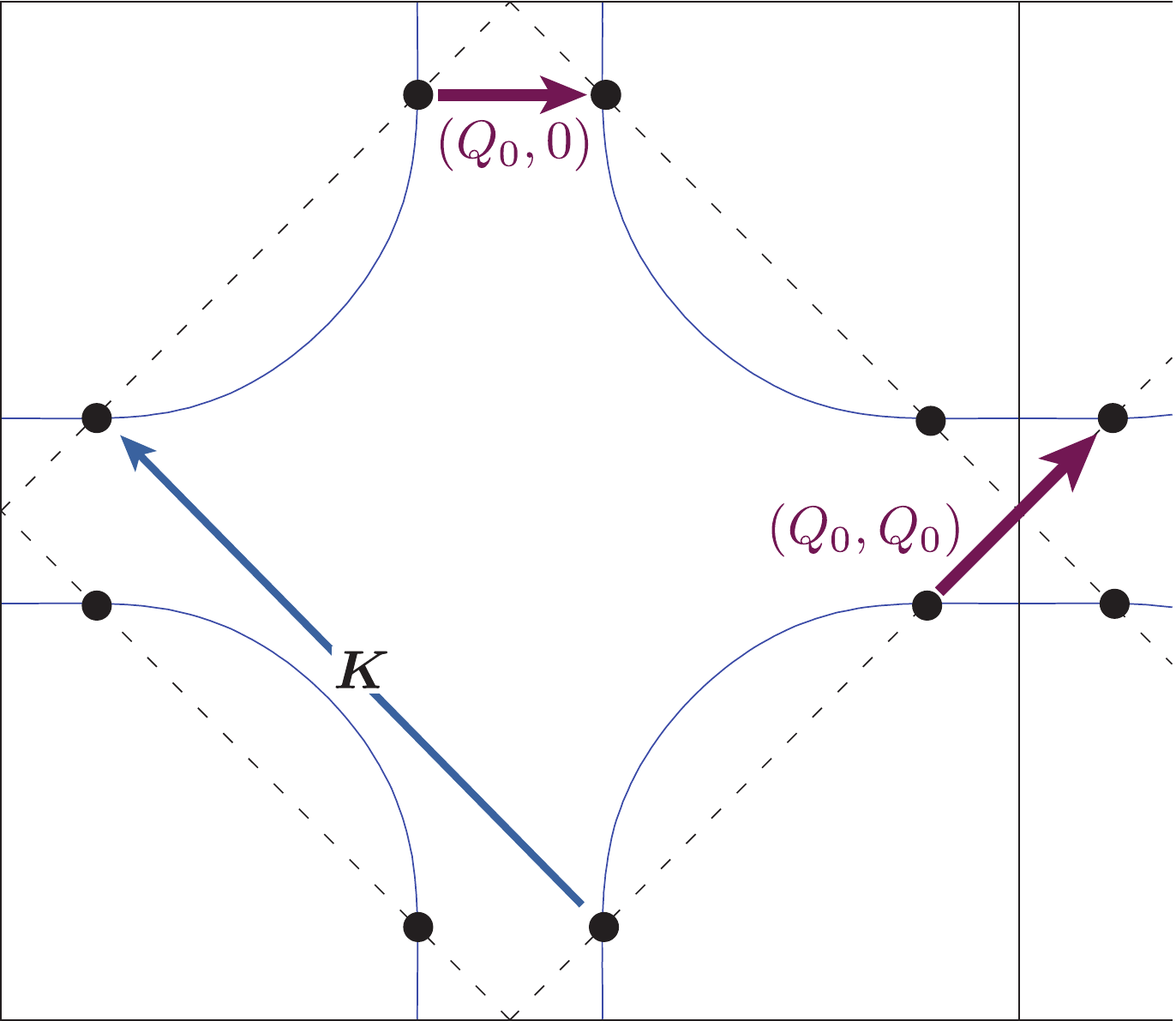}
\caption{Fermi surface with $t_1 = 1$, $t_2 = -0.32$, $t_3=0.128$, and $\mu=-1.11856$.
For this dispersion we have $Q_0 = 4 \pi/11$.
}
\label{fig:bz}
\end{figure}

In our works \cite{paper1,rolando,jay}, ordering wavevectors associated with hot spots on the Fermi surface 
play a special role (see Fig.~\ref{fig:bz}). 
In Section~\ref{sec:hf},  we will introduce the instabilities in the simpler
context of  a `generalized RPA' analysis of a model which includes an on-site
repulsion, $U$, between the electrons. Our main results are in
Section~\ref{sec:sb}, where we will take the limit $U \rightarrow \infty$
using the large $N$ limit of a model with SU($2N$) spin rotation symmetry. In
Section~\ref{sec:dmft} we perform an alternative calculation where the
effective of large repulsion is included via a DMFT self-energy.

\section{RPA analysis}
\label{sec:hf}

This section will carry out a computation similar to that in Ref.~\onlinecite{rolando}, but we will work with a more
general Hamiltonian and use a slightly different formalism. We consider electrons $c_{i \alpha}$ on the sites, $i$, of a square lattice,
with $\alpha = \uparrow, \downarrow$ the spin index, and repeated spin indices, $\alpha,\beta\ldots$, are implicitly summed over.
We work with the following Hamiltonian 
\bea
H &=& H_t + H_C + H_J \nn
H_t &=& -\sum_{i,j} t_{ij} c_{i \alpha}^\dagger c_{j \alpha} -\mu \sum_i c_{i \alpha}^\dagger c_{i \alpha} \nn
H_C &=& U \sum_i c_{i \uparrow}^\dagger c_{i \uparrow} c_{i \downarrow}^\dagger c_{i \downarrow}  
+ \sum_{i<j} V_{ij} c_{i\alpha}^\dagger c_{i \alpha} c_{j \beta}^\dagger c_{j \beta} 
\nn
H_J &=& \sum_{i<j} \sum_a  \frac{J_{ij}}{4} \sigma^a_{\alpha\beta} \sigma^{a}_{\gamma\delta}
c^\dagger_{i\alpha} c_{i \beta} c^\dagger_{j \gamma} c_{j \delta}, \label{ham}
\eea
where $\sigma^a$ are the Pauli matrices with $a=x,y,z$.
We will consider first, second, and third neighbor hopping $t_1$, $t_2$, $t_3$. Similarly, we have first, second,
and third Coulomb and exchange interactions $V_1$, $V_2$, $V_3$ and $J_1$, $J_2$, and $J_3$.

We now introduce our generalized order parameters, $P_{\bQ} (\bk)$ , at wavevector $\bQ$ in the
particle-hole channel by the parameterization
\beq
\left\langle c_{i \alpha}^\dagger c_{j \alpha} \right\rangle = \sum_{\bQ} \left[ \int \frac{d^2 k}{4 \pi^2} P_{\bQ} (\bk) 
e^{i \bk \cdot (\br_i - \br_j )} \right] e^{i \bQ \cdot (\br_i + \br_j)/2}. \label{defP}
\eeq
A conventional charge density wave at 
wavevector $\bQ$ has $P_{\bQ} (\bk)$ independent of $\bk$ so that Eq.~(\ref{defP}) is non-zero
only for $i = j$. However, optimization of the bond energies requires that we allow $P_{\bQ} (\bk )$ to be an arbitrary function
of $\bk$ in the first Brillouin zone. Here, we will find it useful to expand $P_{\bQ} (\bk)$ in terms of a set of orthonormal 
basis functions $\phi_\ell (\bk)$
\beq
P_{\bQ} (\bk ) = \sum_{\ell} \mathcal{P}_{\ell} (\bQ) \phi_\ell (\bk) , \label{defP2}
\eeq
and the coefficients $\mathcal{P}_{\ell} (\bQ)$ become our order parameters.
As we will shortly see, for the Hamiltonians
we work with it is only necessary to include a finite set of values of $\ell$
in Eq.~(\ref{defP2}): we work with the 13 basis functions 
$\phi_\ell (\bk)$ as shown in Table~\ref{tab:basisfunc}.
\begin{table}
  \centering
\begin{tabular}{c|ccc}
$\ell$ & $\phi_\ell(\bk)$           & $\mathcal J_\ell$ & $\mathcal V_\ell$\\
\hline
0      & 1                          & 0                 & $U$ \\
1      & $\cos k_x - \cos k_y$      & $J_1$             & $V_1$\\
2      & $\cos k_x + \cos k_y$      & $J_1$             & $V_1$\\
3      & $2 \sin k_x \sin k_y$      & $J_2$             & $V_2$\\
4      & $2 \cos k_x \cos k_y$      & $J_2$             & $V_2$\\
5      & $\cos(2k_x) - \cos(2k_y)$  & $J_3$             & $V_3$\\
6      & $\cos(2k_x) + \cos(2k_y)$  & $J_3$             & $V_3$
\end{tabular}\hspace{20pt}
\begin{tabular}{c|ccc}
$\ell$ & $\phi_\ell(\bk)$           & $\mathcal J_\ell$ & $\mathcal V_\ell$\\
\hline
\phantom{$J_3$}\\
7      & $\sin k_x - \sin k_y$      & $J_1$             & $V_1$\\
8      & $\sin k_x + \sin k_y$      & $J_1$             & $V_1$\\
9      & $2 \cos k_x \sin k_y$      & $J_2$             & $V_2$\\
10     & $2 \sin k_x \cos k_y$      & $J_2$             & $V_2$\\
11     & $\sin(2k_x) - \sin(2k_y)$  & $J_3$             & $V_3$\\
12     & $\sin(2k_x) + \sin(2k_y)$  & $J_3$             & $V_3$\\
\end{tabular}
  \caption{Relevant basis functions}
  \label{tab:basisfunc}
\end{table}


We take the index $\ell = 0,1,\dots 12$. Note that the orderings with $\ell = 0, \ldots 6$ represent charge/bond density waves
which preserve time-reversal, while those with $\ell = 7, \ldots 12$ represent states with spontaneous currents which break 
time-reversal.

A key step is to rewrite the interaction terms Eq.~(\ref{ham}) in the following form
\bea
H_J + H_C &=&  \sum_{\bk, \bk', \bq} \sum_{\ell = 0}^{12} \phi_\ell (\bk) \phi_\ell (\bk') \left[ 
\sum_a \frac{\mathcal{J}_\ell}{8} c^\dagger_{\bk' - \bq/2,\alpha} \, \sigma^a_{\alpha\beta} \,c_{\bk - \bq/2, \beta} \, c^\dagger_{\bk + \bq/2,\gamma} \, \sigma^a_{\gamma\delta} \, c_{\bk' + \bq/2, \delta} \right . \nn
&~&\quad \quad \left.
+ \frac{\mathcal{V}_\ell}{2} c^\dagger_{\bk' - \bq/2,\alpha}  \,c_{\bk - \bq/2, \alpha} \, c^\dagger_{\bk + \bq/2,\beta} \,  c_{\bk' + \bq/2, \beta} \right] \label{Hbasis}
\eea
where the $\phi_\ell (\bk)$ are 13 orthonormal basis functions in Table~\ref{tab:basisfunc}, and $\mathcal{J}_\ell$ and $\mathcal{V}_\ell$ are 
the corresponding couplings shown in Table~\ref{tab:basisfunc}.
The appearance of a finite set of basis functions in Eq.~(\ref{Hbasis}) is the
reason we are able to truncate the expansion in Eq.~(\ref{defP2}). 

\begin{figure}
\includegraphics[width=6.5in]{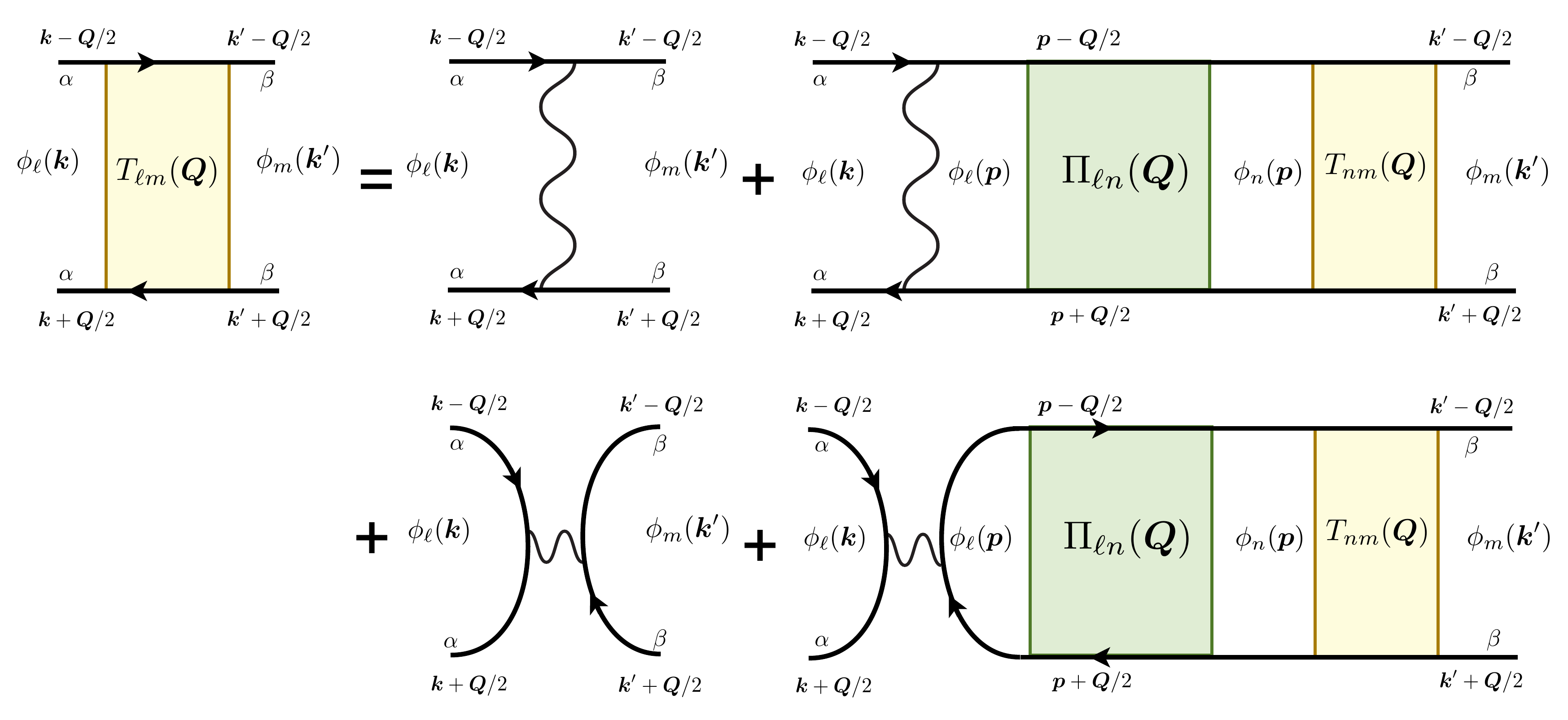}
\caption{Schematic equation for the $T$-matrix in the spin-singlet particle-hole channel with total momentum $\bQ$}
\label{fig:tmatrix}
\end{figure}

We can now use the basis $\phi_\ell (\bk)$ to also decompose 
 the Bethe-Salpeter equation in the spin-singlet, particle-hole channel, as shown in Fig.~\ref{fig:tmatrix}. 
The eigenmodes of the resulting $T$-matrix
$T_{\ell m} (\bQ)$ will determine the structure of the ordering, $\mathcal{P}_\ell (\bQ)$ at the wavevector $\bQ$.
 
Summing ladder diagrams for both direct and exchange interactions we obtain
\bea
T_{\ell m} (\bQ) &=& \left(\frac{3}{4} \mathcal{J}_\ell  + \mathcal{V}_\ell  \right) \delta_{\ell m} 
- 2\delta_{\ell,0} \delta_{m,0} W (\bQ)   \label{teqn}
 \\
&+& \frac{1}{2} \sum_{n=0}^{12}
\left(\frac{3}{4} \mathcal{J}_\ell + \mathcal{V}_\ell  \right) 
 \Pi_{\ell n} (\bQ) T_{nm} (\bQ)
 - \delta_{\ell,0} \sum_{n=0}^{12} W (\bQ) \Pi_{0n} (\bQ) T_{nm} (\bQ) \nonumber
\eea
where
\beq
W (\bQ) \equiv \sum_{\ell=0}^{12} \mathcal{V}_{\ell} \phi_{\ell} (0) \phi_{\ell} (\bQ) 
\eeq
is the direct interaction, and 
$\Pi_{\ell m} (\bQ)$ is a $13\times 13$ matrix which 
is the polarizability of the Hamiltonian $H_C$
\beq
\Pi_{\ell m} (\bQ) = 2 \sum_{\bk} \phi_\ell (\bk) \phi_m (\bk) \frac{ f(\varepsilon(\bk - \bQ/2)) - f(\varepsilon(\bk + \bQ/2))}{
\varepsilon(\bk +\bQ/2)- \varepsilon(\bk - \bQ/2)} \label{defPilm}
\eeq
with $\varepsilon(\bk)$ is the single particle dispersion:
\beq
\varepsilon (\bk) = -2t_1 (\cos (k_x) + \cos(k_y)) - 4 t_2 \cos (k_x) \cos (k_y) -  2t_3 (\cos (2k_x) + \cos(2k_y)) - \mu.
\eeq
We choose the dispersion $\varepsilon (\bk)$ to have hot spots which intersect the magnetic Brillouin zone boundary,
as shown in Fig.~\ref{fig:bz}.
The hot spots for this dispersion are separated by the vectors shown with $Q_0 = 4\pi/11$.
Note that $Q_0$ is simply a geometric property of the Fermi surface, and plays no special role in the Hamiltonian.

\begin{figure}
\includegraphics[width=4in]{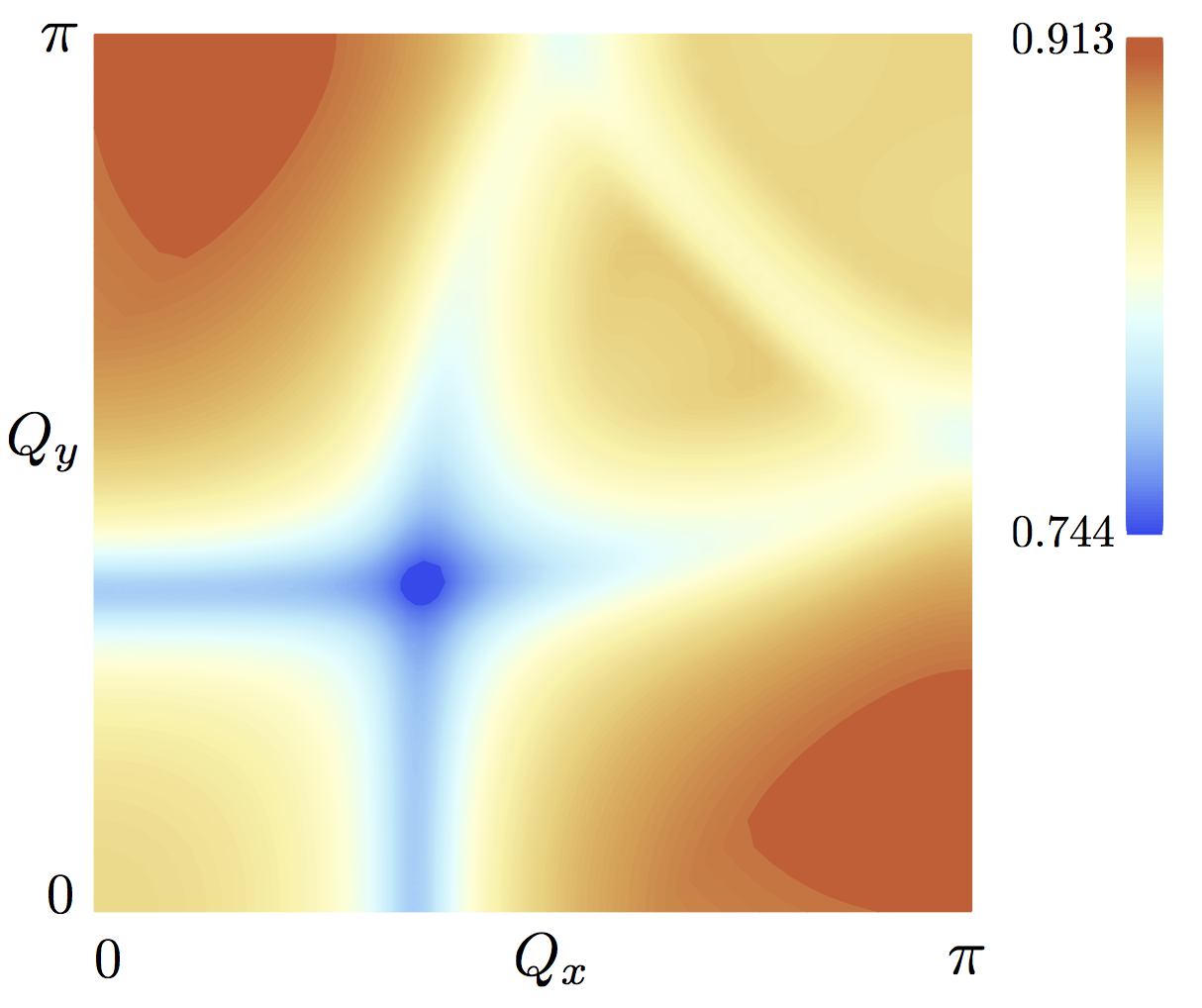}
\caption{Lowest eigenvalues, $\lambda_\bQ$, of the 13$\times$13 matrix in Eq.~(\ref{lambdaQ}) at
a temperature $T=0.06$.
The Fermi surface is as in Fig.~\ref{fig:bz}, and the interaction couplings are $J_1 = 0.5$, $J_2 = 0.2$, $J_3=0.05$,
$U=0$, $V_1=0$, $V_2=0$, $V_3=0$.
Minimized over $\bQ$, the lowest eigenvalue is at $\bQ = (0.38, 0.38)\pi$; this is very close to the value
$Q_0 = 0.36\pi$ as determined from the Fermi surface in Fig.~\ref{fig:bz}. The eigenvector at $\bQ =(0.38, 0.38)\pi$
is $P_{\bQ} (\bk) = 0.9996 (\cos (k_x) - \cos(k_y)) + 0.0275 (\cos(2 k_x) - \cos (2 k_y))$.
}
\label{fig:hf12}
\end{figure}

\begin{figure}
\includegraphics[width=4in]{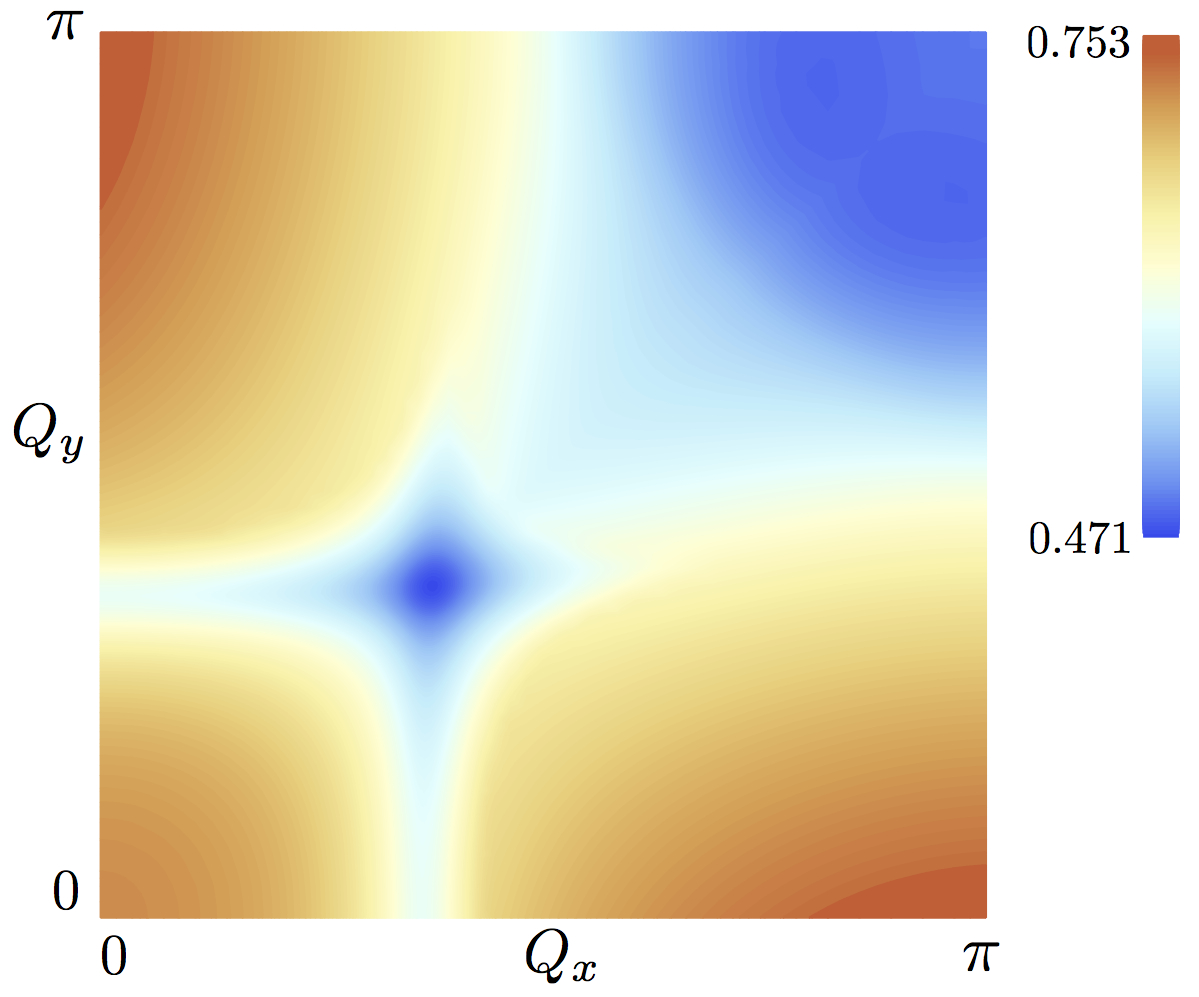}
\caption{As in Fig.~\ref{fig:hf12}, with all parameters the same apart from $U=1$, $V_1 = 0.4$, $V_2=0.2$,
and $V_3 = 0.05$.
Minimized over $\bQ$, the lowest eigenvalue is again at $\bQ = (0.38, 0.38)\pi$ and the corresponding eigenvector
is $P_{\bQ} (\bk) = 0.9995 (\cos (k_x) - \cos(k_y)) + 0.0312 (\cos(2 k_x) - \cos (2 k_y))$. Now there are also small, but slightly larger, eigenvalues
near $\bQ = (\pi, \pi)$ with eigenvectors which break time-reversal.
}
\label{fig:hf13}
\end{figure}

By rearranging terms in Eq.~(\ref{teqn}), we see that the charge-ordering instability is determined by the lowest eigenvalues, $\lambda_\bQ$ of the matrix
\beq
\delta_{\ell m} - \frac{1}{2}
\left(\frac{3}{4} \mathcal{J}_\ell + \mathcal{V}_\ell  \right) 
 \Pi_{\ell m}(\bQ)  + \delta_{\ell,0} W (\bQ) \Pi_{0m} (\bQ), \label{lambdaQ}
\eeq
and the $\mathcal{P}_{m} (\bQ)$ are determined by the corresponding right eigenvector.
The values of $\lambda_\bQ$ are shown in Figs.~\ref{fig:hf12} and~\ref{fig:hf13} for the metallic state
with the Fermi surface in Fig.~\ref{fig:bz}.

In Fig.~\ref{fig:hf12} we consider a case with vanishing on-site interactions, as in Ref.~\onlinecite{rolando}. 
As found previously, the lowest eigenvalue is at $\bQ \approx (Q_0, Q_0)$ and the corresponding eigenvector
is purely $d$-wave. 

\begin{figure}
\begin{center}
\includegraphics[width=3in]{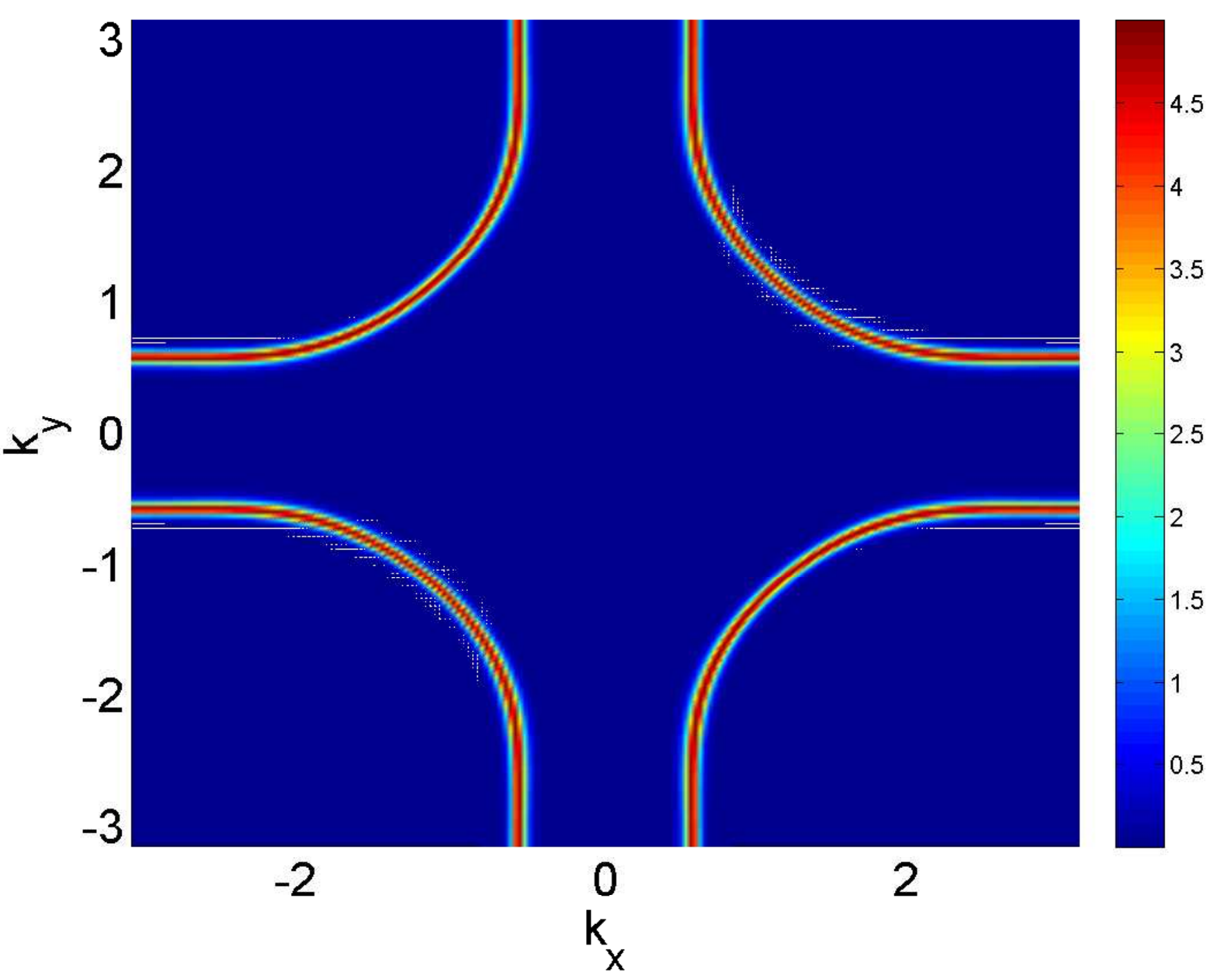}
\includegraphics[width=3in]{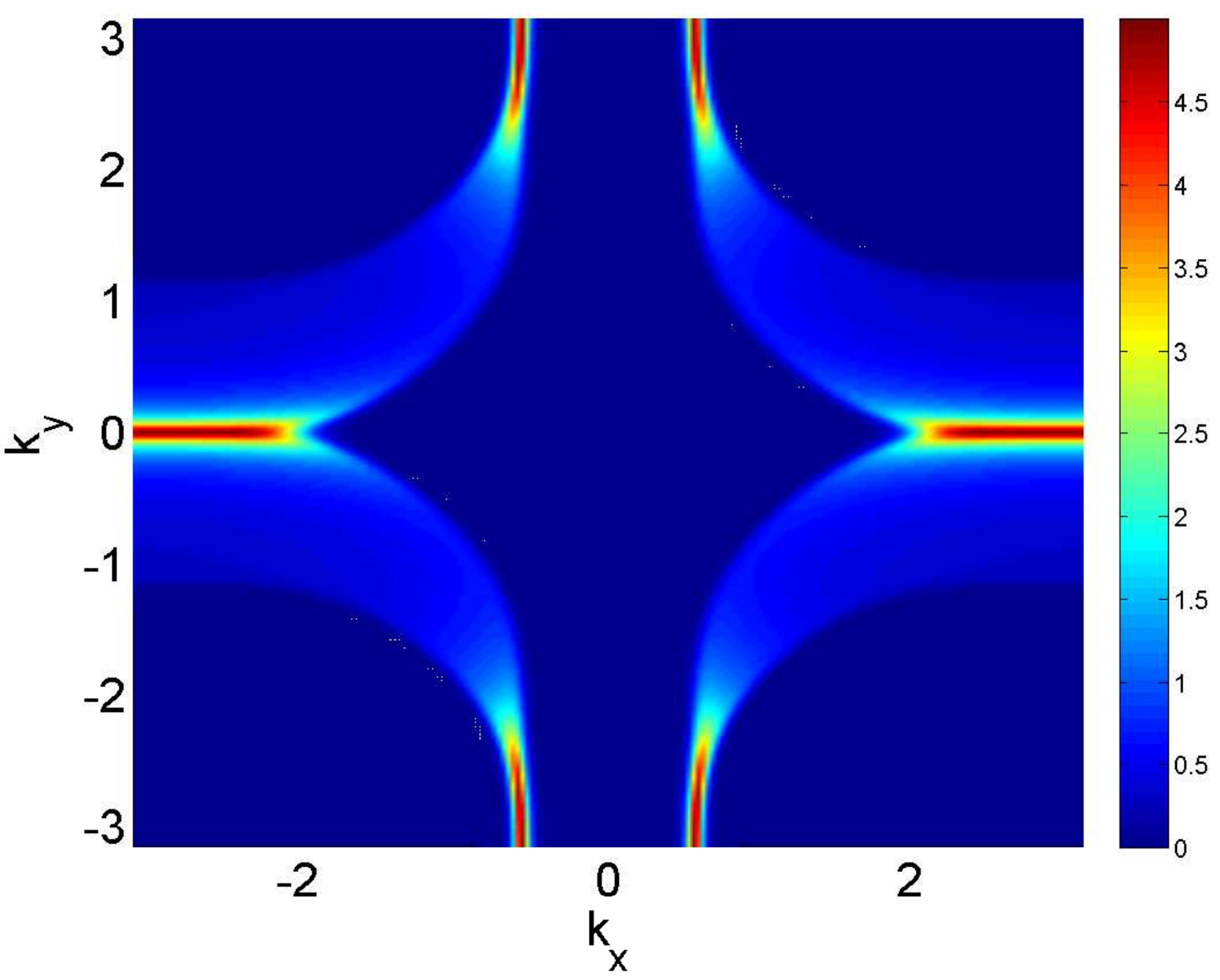}\\
\includegraphics[width=3in]{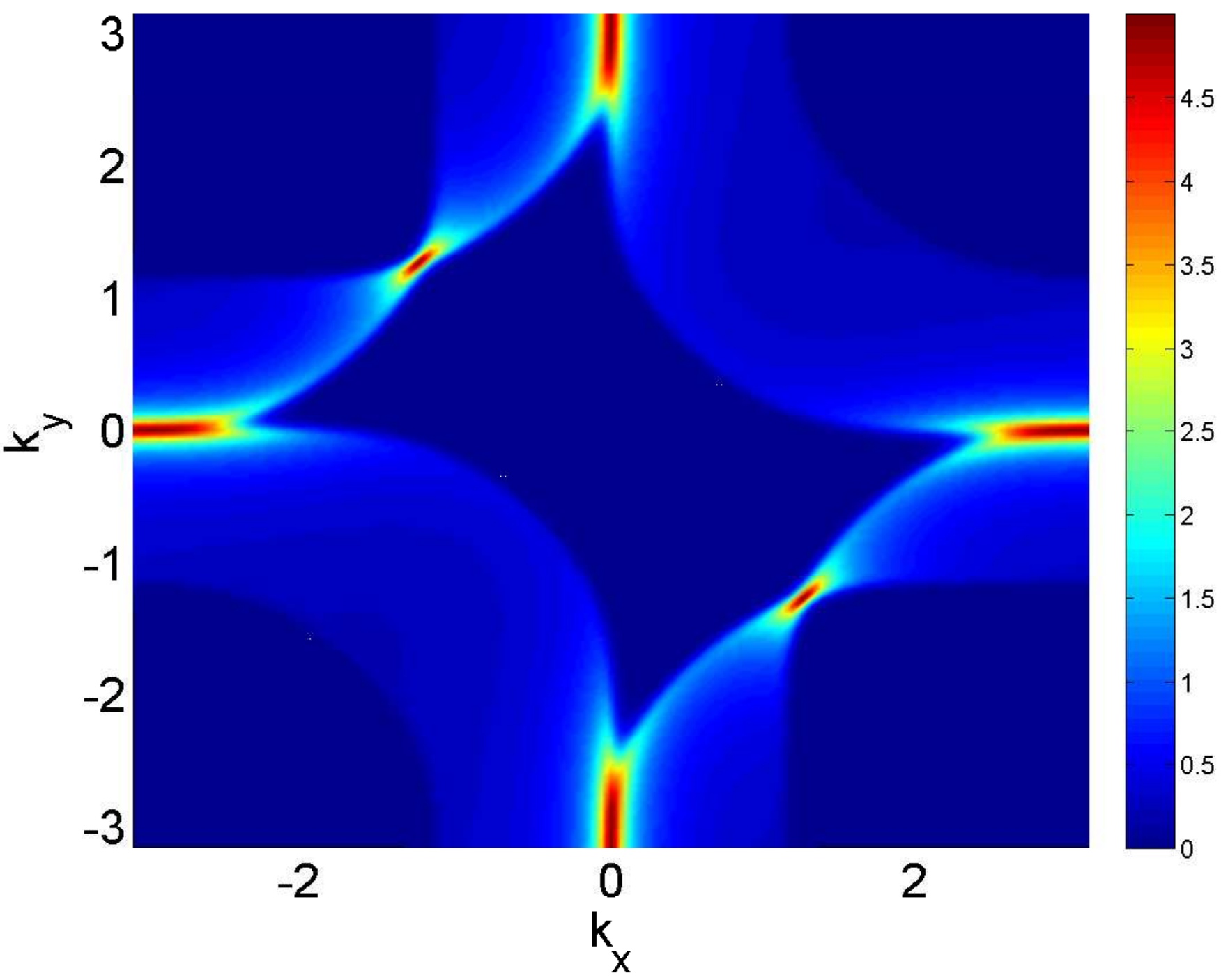}
\includegraphics[width=3in]{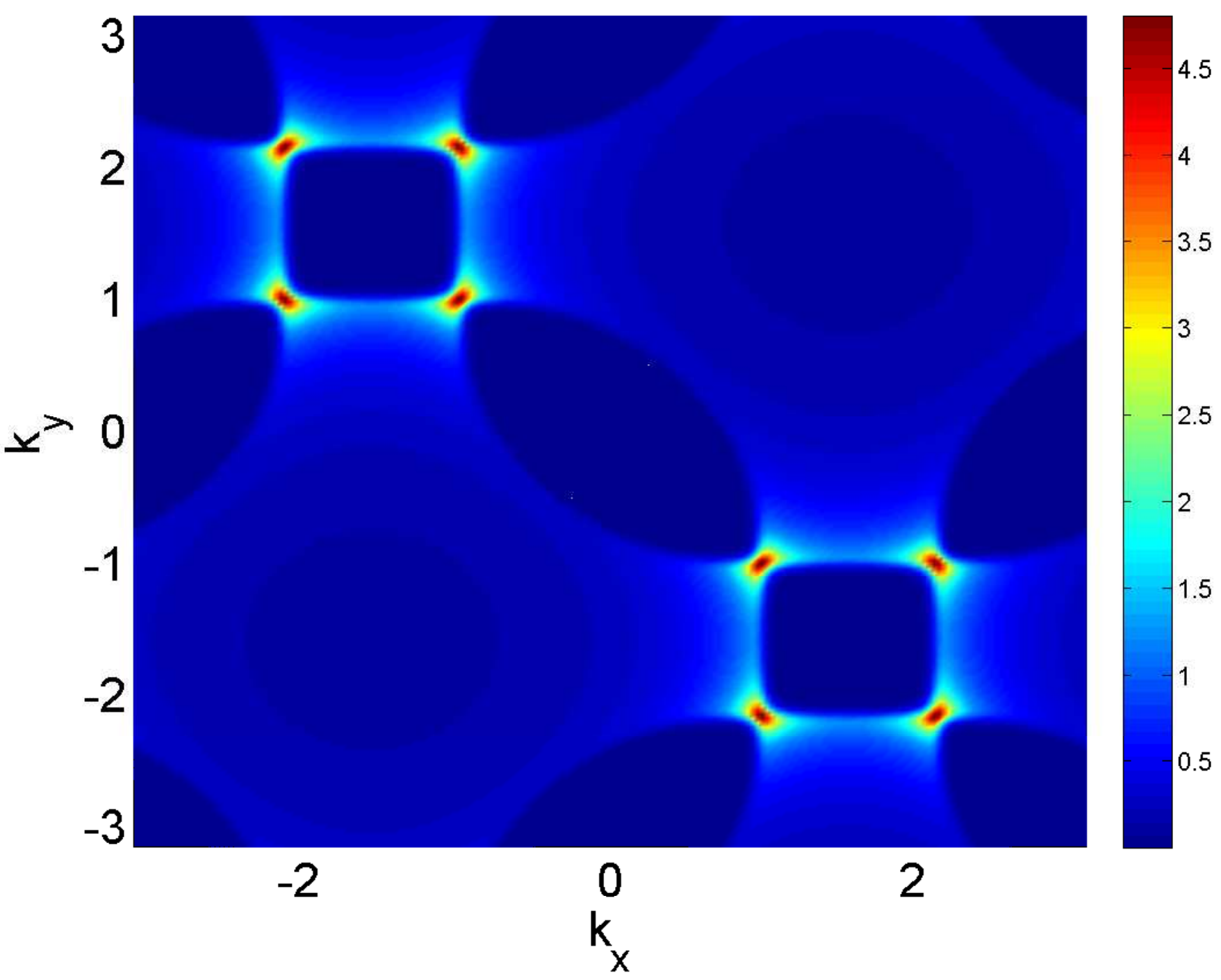}
\end{center}
\caption{Color plots of the magnitude of the integrand $\Pi (\bk, \bQ)$ in
  Eq.~(\ref{eq:Pikq}) for $T=0.05$ and hopping parameters as before as function of
  $\bk$ for different   $\bQ=0$, $Q_0 (1,0)$, $Q_0 (1,1)$, $(\pi,\pi)$ (from
  top left to bottom right). The result
  for $\bQ=0$ is strongly peaked at the Fermi surface. We see that for
  $\bQ=Q_0 (1,0), Q_0 (1,1)$ we obtain large matrix elements $\Pi_{11} (\bQ)$ with $d$-wave
  symmetry $\phi_1(\bk)$ as $\phi_1(\bk)^2$ is peaked at $\pm (\pi,0)$, $\pm (0,\pi)$. For $\bQ=
  (\pi,\pi) $ the largest contribution is for $\Pi_{77} (\bQ)$ with the
  $\phi_7(\bk)$ eigenfunction where $\phi_7(\bk)^2$ is peaked at $\pm
  (-\pi/2,\pi/2)$.}
\label{fig:Pikq}
\end{figure}
We turn on Coulomb interactions in Fig.~\ref{fig:hf13}, while keeping other parameters the same.
The main change is that the eigenvalues near $\bQ = (\pi, \pi)$ become significantly smaller.
The eigenvectors in this region of $\bQ$ break time-reversal \cite{rolando}, and the eigenvector 
at $\bQ = (\pi, \pi)$ is $P_\bQ (\bk) = \sin (k_x) - \sin (k_y)$. Some
intuition about which wavevector is favored with the corresponding eigenvector
can be gained from the plots of the relevant integrand in the instability equation.
\beq
\Pi (\bk,\bQ) = \frac{ f(\varepsilon(\bk - \bQ/2)) - f(\varepsilon(\bk + \bQ/2))}{
\varepsilon(\bk +\bQ/2)- \varepsilon(\bk - \bQ/2)} \label{eq:Pikq}
\eeq
in Fig.~\ref{fig:Pikq}. 


In both Figs.~\ref{fig:hf12} and~\ref{fig:hf13}, there is a ridge of minima extending from $(Q_0,Q_0)$
to $(0, Q_0)$, and also to $(Q_0,0)$. The latter wavevectors are close to the
experimentally observed values.\cite{comin}
At the wavevector $\bQ = (0, Q_0)$, the charge ordering eigenvector for Fig.~\ref{fig:hf13} is
\bea
P_\bQ (\bk) &=& -0.352 -0.931 \bigl[ \cos (k_x) - \cos(k_y) \bigr] + 0.017 \bigl[\cos (k_x) + \cos (k_y)\bigr] 
\label{splusd} \\
&-& 0.168 \cos (k_x) \cos(k_y)
 - 0.028 \bigl[ \cos (2k_x) - \cos(2k_y) \bigr] + 0.029 \bigl[ \cos (2k_x) + \cos(2k_y) \bigr]. \nonumber
\eea
So the largest component at this $\bQ$ remains a $d$-wave on the nearest neighbor bonds, 
but now there is a significant on-site
density wave.

There is also a local minimum in Fig.~\ref{fig:hf13} at $\bQ = (\pi, \pi)$. Here the eigenvector is 
\beq
P_\bQ (\bk) = \sin (k_x) - \sin (k_y) . \label{sfP}
\eeq
This represents the ``staggered flux'' state of Refs.~\onlinecite{marston,kotliar,sudip,leewen,laughlin}.
This state was called a ``$d$-density wave'' in Ref.~\onlinecite{sudip}, which is an unfortunate terminology
from our perspective. With our identification of the bond expectation values in Eq.~(\ref{defP}), this state is actually
a $p$-density wave,\cite{rolando} as is evident from Eq.~(\ref{sfP}).

\section{$U \rightarrow \infty$ limit}
\label{sec:sb}

We will continue to work with the Hamiltonian in Eq.~(\ref{ham}), but will now set $U=\infty$.
The $U=\infty$ constraint is implemented by the auxiliary-boson decomposition
\beq
c_{i \alpha} = b_i^\dagger f_{i \alpha}
\eeq
where $b_i$ is a canonical boson and $f_{i \alpha}$ is a canonical fermion, along with the constraint
\beq
b_i^\dagger b_i + f_{i \alpha}^\dagger f_{i \alpha} = N. \label{const}
\eeq
Here we allow the index $\alpha = 1 \ldots 2N$, so that the model has 
SU($2N$) symmetry. The constraint can then be systematically implemented in the large $N$ limit.\cite{marston,grilli}

We can write the SU($2N$) Lagrangian as
\bea
\mathcal{L}
 &=& \sum_i \left[ f_{i \alpha}^\dagger \left( \frac{\partial}{\partial \tau} - \mu + i \lambda_i \right) f_{i \alpha} 
+  b_i^\dagger \left( \frac{\partial}{\partial \tau} + i \lambda_i \right) b_i - i N \lambda_i \right]
 \nn
&~& - \frac{1}{N} \sum_{i,j} t^{0}_{ij} b_i b_j^\dagger  f_{i \alpha}^\dagger f_{j \alpha}  
+ \frac{1}{N} \sum_{i<j} V_{ij} (N - b_i^\dagger b_i) (N - b_j^\dagger b_j) \nn
&~&  + \sum_{i<j} J_{ij} \left( 2 N |P_{ij}|^2  - P_{ij}^\ast f_{i \alpha}^\dagger f_{j \alpha} - P_{ij} f_{j \alpha}^\dagger f_{i \alpha}
\right) \label{L}
\eea
where we have decoupled the exchange interaction by a Hubbard-Stratanovich variable $P_{ij}$ 
residing on the bonds, and 
absorbed a contribution of $-J_{ij}/4$ into the definition of $V_{ij}$. Also, we have written the fermion hopping as
$t^{0}$ because this will undergo a renormalization before determining the fermion dispersion.

\subsection{$N=\infty$ theory}

We take $b_i = \sqrt{N}\, \overline{b}$, $ \lambda_i = -i \overline{\lambda}$ and $P_{ij} = \overline{P}_{1,2,3}$ for $ij$ 
first, second, third neighbors.
Then the fermion dispersion is
\beq
H_{f} = \sum_{\bk} E (\bk) f_{\bk \alpha}^\dagger f_{\bk \alpha}
\eeq
with
\beq
E (\bk) = - \overline{b}^2 \gamma (\bk ) - \gamma_J (\bk) - \mu +
\overline{\lambda} ,
\eeq
where 
\beq
\gamma (\bk) =  2 t^{0}_1 ( \cos(k_x) +  \cos(k_y)) + 4 t^{0}_2 \cos (k_x) \cos (k_y) + 
2 t^{0}_3 ( \cos( 2 k_x) +  \cos(2 k_y)) ,
\eeq
and
\beq
\gamma_J (\bk) =  2 J_1 P_1 ( \cos(k_x) +  \cos(k_y)) + 4 J_2 P_2 \cos (k_x)
\cos (k_y) + 2 J_3 P_3 ( \cos( 2 k_x) +  \cos(2 k_y)) .
\eeq
From these relations we see that the renormalized fermion hopping parameters are 
\beq 
t_i = t^{0}_i \overline{b}^2 + J_i P_i,
\eeq
where $i=1,2,3$.

The mean-field equations for the $P$'s are obtained from the $N=\infty$ saddle point condition, which yield
\bea
P_1 &=& \sum_\bk \cos (k_x) \, f (E(\bk)) \nn
P_2 &=& \sum_\bk \cos (k_x + k_y) \, f (E(\bk)) \nn
P_3 &=& \sum_\bk \cos (2 k_x) \, f (E(\bk)) .
\eea
The constraint equation from the saddle point of $\lambda_i$ is
\beq
\overline{b}^2 = 1  -  2 \sum_\bk f(E (\bk)) .
\eeq
And finally, the saddle point equation for $\overline{b}$ is
\beq
\overline{\lambda} = 4 (V_1 + V_2 + V_3) ( 1- \overline{b}^2) + 2 \sum_\bk
\gamma( \bk) f(E (\bk)) .
\eeq

\subsection{$1/N$ fluctuations}

It is useful to manipulate the exchange interactions into the following form
\bea 
H_J &=& -\sum_{i,j} \frac{J_{ij}}{4 N} f_{i \alpha}^\dagger f_{j \alpha} f_{j \beta}^\dagger f_{i \beta} \nn
&=&- \frac{1}{4 N} \sum_{\bk, \bk', \bQ} \left( \sum_{a} J_{i,i+a} e^{i (\bk - \bk') \cdot {\bm a}} \right) f_{\bk + \bQ/2, \alpha}^\dagger f_{\bk - \bQ/2,\alpha} 
f_{\bk' - \bQ/2, \beta}^\dagger f_{\bk' - \bQ/2,\beta} \nn
&=& -\frac{1}{4N} \sum_\bQ \sum_{\ell=1}^{12} \mathcal{J}_\ell \left|
  \sum_{\bk} \phi_\ell (\bk) f_{\bk + \bQ/2, \alpha}^\dagger f_{\bk -
    \bQ/2,\alpha} \right|^2 ,
\eea
where $a$ extends over first, second, and third neighbors, and
the $\mathcal{J}_\ell$ and the $\phi_\ell$ are the same as in Table~\ref{tab:basisfunc}. Note that in this section the index $\ell$ extends
from $\ell = 1$ to $\ell = 12$ (implicitly, where not noted), and the $\ell=0$ basis states in Table~\ref{tab:basisfunc} are not included.
Now we can decouple the exchange coupling to 
\beq
H_J = \sum_{\bQ} \sum_{\ell=1}^{12} \mathcal{J}_\ell \left[ N
  |\mathcal{P}_{\ell} (\bQ) |^2 - \sum_{\bk} \mathcal{P}_\ell (-\bQ) \phi_\ell
  (\bk) f_{\bk + \bQ/2, \alpha}^\dagger f_{\bk - \bQ/2,\alpha} \right] ,
\eeq
with $ \mathcal{P}_\ell (-\bQ) = \mathcal{P}_{\ell}^\ast (\bQ)$.
We can now see that the $\mathcal{P}_{\ell} (\bQ)$ are similar to the order parameters as those introduced in Eq.~(\ref{defP2}),
but they now refer to the fermions $f_\alpha$ rather than the electrons $c_\alpha$. These differ by a factor of $\overline{b}$
in the large $N$ limit, and so the corresponding $\mathcal{P}_\ell (\bQ)$ differ by a factor of $\overline{b}^2$. 
The mean-field values of the $\mathcal{P}_\ell (\bQ)$ are
\beq
\overline{\mathcal{P}_\ell} (\bQ) =  \delta_{\bQ,0} \, \{0,2P_1,0,2P_2,0,2P_3,0,0,0,0,0,0\}.
\eeq 

For the fluctuations about mean-field, we fix the unitary gauge, and work at zero frequency of all bosonic fields.
Then we can parameterize the fluctuations as
\bea
\mathcal{P}_\ell (\bQ) &=& \overline{\mathcal{P}_\ell} (\bQ) +
\frac{1}{\sqrt{\mathcal{J}_\ell}} p_\ell (\bQ) , \\
\lambda_{i} &=& -i \overline{\lambda} + \sum_\bQ \lambda (\bQ) e^{i \bQ \cdot
  \br_i } ,\\
b_{i} &=& \sqrt{N} \, \overline{b} + \sqrt{N} \sum_\bQ b (\bQ) e^{i \bQ \cdot
  \br_i } ,
\eea
where $\lambda (-\bQ) = \lambda^\ast (\bQ)$, $b (-\bQ) = b^\ast (\bQ)$, $p_\ell (-\bQ) = p_\ell^\ast (\bQ)$.
Then the Lagrangian (\ref{L}) can be written as
\bea
\mathcal{L} &=& \mathcal{L}_0 + N \sum_{\bQ} \Biggl[ \sum_\ell |p_\ell (\bQ) |^2 +  2 i \overline{b} \sum_{\bQ} \lambda (\bQ) b(-\bQ) \\
&+& \left[ \overline{\lambda} + 4(V_1+V_2+V_3)(\overline{b}^2 -1)  + 2 \overline{b}^2 \gamma_V (\bQ ) \right] b (\bQ) b(-\bQ) \Biggr] \nn
&+& \sum_{\bk}  f_{\bk \alpha}^\dagger \left( \frac{\partial}{\partial \tau} + E (\bk) \right) f_{\bk \alpha}
- \overline{b} \sum_{\bk, \bQ}\left[ \gamma (\bk - \bQ/2) + \gamma(\bk + \bQ/2) \right] b(\bQ) f_{\bk+\bQ/2, \alpha}^\dagger f_{\bk -\bQ/2,\alpha}
\nn
& -&  \sum_{\bk, \bQ_1,\bQ_2} \gamma (\bk) b (\bQ_1) b (\bQ_2) f_{\bk+\bQ_1,
  \alpha}^\dagger f_{\bk -\bQ_2,\alpha} +  \sum_{\bk, \bQ} \left[i \lambda
  (\bQ) - \sqrt{\mathcal{J}_\ell} \, p_\ell (-\bQ) \, \phi_\ell (\bk) \right]
f_{\bk+\bQ/2, \alpha}^\dagger f_{\bk -\bQ/2,\alpha}  ,
\nonumber
\eea
where
\beq
\gamma_V (\bk) =  2 V_1 ( \cos(k_x) +  \cos(k_y)) + 4 V_2 \cos (k_x) \cos
(k_y) + 2 V_3 ( \cos( 2 k_x) +  \cos(2 k_y)) .
\eeq
We integrate out the fermions and obtain
\beq
\mathcal{L} = \mathcal{L}_0 + \frac{N}{2} \sum_{\bQ} \biggl[  \left(p_\ell (-\bQ), b(-\bQ), \lambda(-\bQ)  \right)
\left(\begin{array}{ccc}
2 \delta_{\ell m} - \sqrt{\mathcal{J}_\ell \mathcal{J}_m} \, \Pi_{\ell m} (\bQ) & K_{4\ell} (\bQ) &K_{5\ell} (\bQ) \\
K_{4 m} (\bQ) & K_1 (\bQ) & K_2 (\bQ) \\
K_{5m} (\bQ) & K_2 (\bQ) & K_3 (\bQ)  
\end{array}\right)
\left(\begin{array}{c}
 p_{m} (\bQ) \\ b(\bQ) \\ \lambda (\bQ)  \end{array}\right), \label{LP}
\eeq
where
\bea
K_1 (\bQ) &=&  2 \overline{\lambda} + 8(V_1+V_2+V_3) (\overline{b}^2 - 1) + 4
\overline{b}^2 \gamma_V (\bk ) ,\nn 
&~& + 
\sum_{\bk} \left[ -4 \gamma(\bk) f( E (\bk + \bQ)) - \overline{b}^2\left[
    \gamma (\bk - \bQ/2) + \gamma(\bk + \bQ/2) \right]^2 \Pi (\bk, \bQ) \right] ,\nn
K_2 (\bQ) &=& 2i \overline{b} + i \overline{b} \sum_{\bk}  \left[ \gamma (\bk
  - \bQ/2) + \gamma(\bk + \bQ/2) \right] \Pi (\bk, \bQ) , \nn 
K_3 (\bQ) &=&  \sum_{\bk} \Pi (\bk, \bQ) , \nn
K_{4\ell} (\bQ) &=& - \overline{b} \sqrt{\mathcal{J}_\ell} \sum_\bk  \phi_\ell
(\bk)  \left[ \gamma (\bk - \bQ/2) + \gamma(\bk + \bQ/2) \right]  \Pi (\bk,
\bQ) ,\nn 
K_{5\ell} (\bQ) &=& i \sqrt{\mathcal{J}_\ell}  \sum_{\bk} \phi_{\ell} (\bk)
\Pi (\bk, \bQ) , 
\eea
with
\begin{equation}
\Pi (\bk, \bQ) = 2 \frac{ f(E(\bk - \bQ/2)) - f (E(\bk +
  \bQ/2))}{E(\bk + \bQ/2)) - E(\bk - \bQ/2)},
\label{eq:pikq}
\end{equation}
and $\Pi_{\ell m} (\bQ)$ defined as in Eq.~(\ref{defPilm}). 

\begin{figure}
\includegraphics[width=4in]{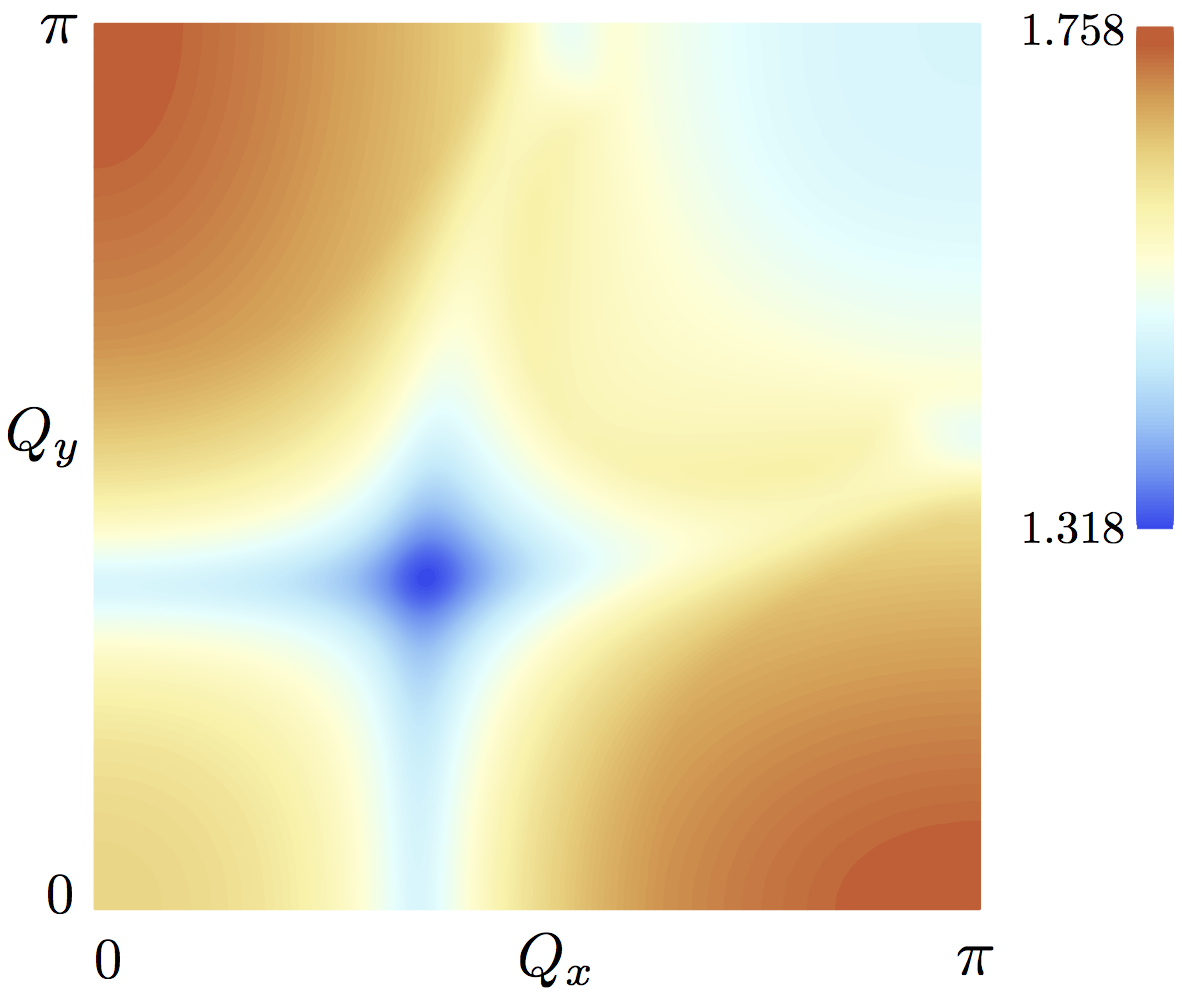}
\caption{As in Fig.~\ref{fig:hf13}, with all parameters the same apart from the $U=\infty$ limit taken via the auxiliary-boson method.
As noted in the text, the present eigenvalues have to be compared with twice the eigenvalues in Fig.~\ref{fig:hf13}.
The structure of the eigenvalues and eigenvectors is very similar to Fig.~\ref{fig:hf13}, with the main difference that the strength
of the sub-dominant instability to the time-reversal symmetry breaking staggered flux state near $(\pi, \pi)$ is now weaker.}
\label{fig:largensb}
\end{figure}
We now perform the Gaussian integrals over the fields $\lambda (\bQ)$ and $b (\bQ)$, and then 
diagonalize the resulting quadratic form for the fields $p_\ell (\bQ)$. 
This step is the analog of our solution of the Bethe-Salpeter
equation in Section~\ref{sec:hf}. Note that the quadratic form for the $p_\ell (\bQ)$ in Eq.~(\ref{LP}) begins with a 
$2 \delta_{\ell m}$, which is to be compared with the $\delta_{\ell m}$ in Eq.~(\ref{lambdaQ}); consequently, the present eigenvalues $\lambda_\bQ$ are to be compared with {\em twice\/} the eigenvalues in Section~\ref{sec:hf}. We also note
that a related computation was carried out in a different gauge in the early work of Ref.~\onlinecite{wang},
but they did not consider Fermi surfaces with hot spots.

Our results for the $\lambda_\bQ$ are shown in Fig.~\ref{fig:largensb}, with the same set of parameters
as in Fig.~\ref{fig:hf13} in Section~\ref{sec:hf} but with the $U=\infty$ limit taken in the large $N$ method.
The results are very similar, but the eigenvalues of the time-reversal symmetry breaking 
`staggered flux' state near $\bQ = (\pi, \pi)$ are a bit larger now.
The global minimum of $\lambda_\bQ$ remains at $\bQ = (0.38, 0.38) \pi$ and the corresponding eigenvector
is purely $d$ wave (note that the values of $\ell$ extend over $1 \ldots 12$):
\beq
p_\ell (\bQ) = \{0.996,\, 0,\, 0,\, 0,\, 0.087,\, 0,\,
 0,\, 0,\, 0,\, 0,\, 0,\, 0\} \quad, \quad \bQ= (0.38, 0.38)\pi\,.
\eeq
For the local minimum at $\bQ = (0, 0.38) \pi$ the eigenvector is a mixture of $s$ and $d$ wave, as in Eq.~(\ref{splusd}):
\beq
p_\ell (\bQ) = \{ 
0.988, \, 0.001,\, 0,\, 0.112,\, 0.077,\, -0.079, 0,\, 0,\, 0,\, 0,\, 0,\, 0
\} \quad, \quad \bQ= (0, 0.38)\pi\,. \label{splusd2}
\eeq

\begin{figure}
\includegraphics[width=3in]{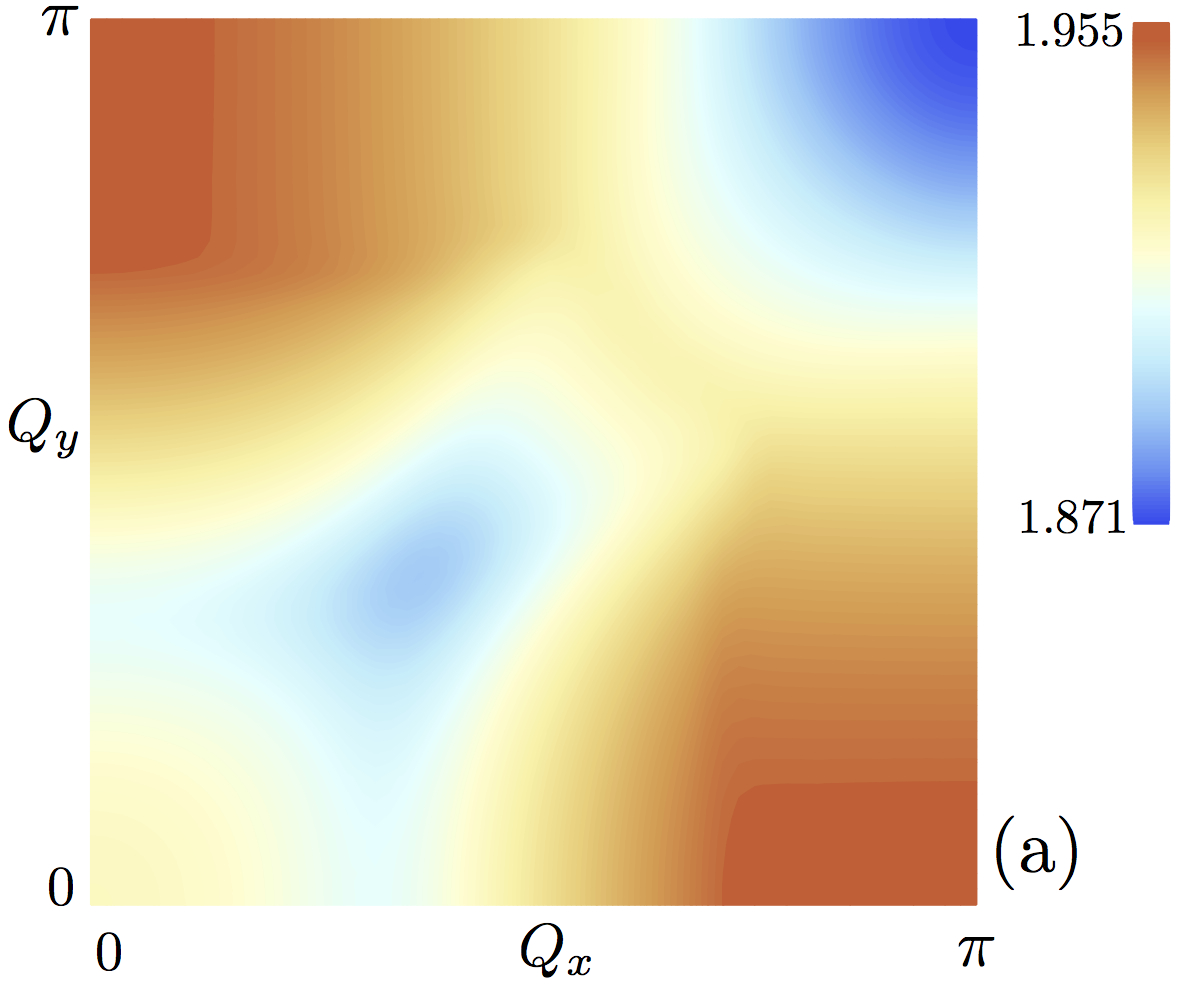}
\includegraphics[width=3in]{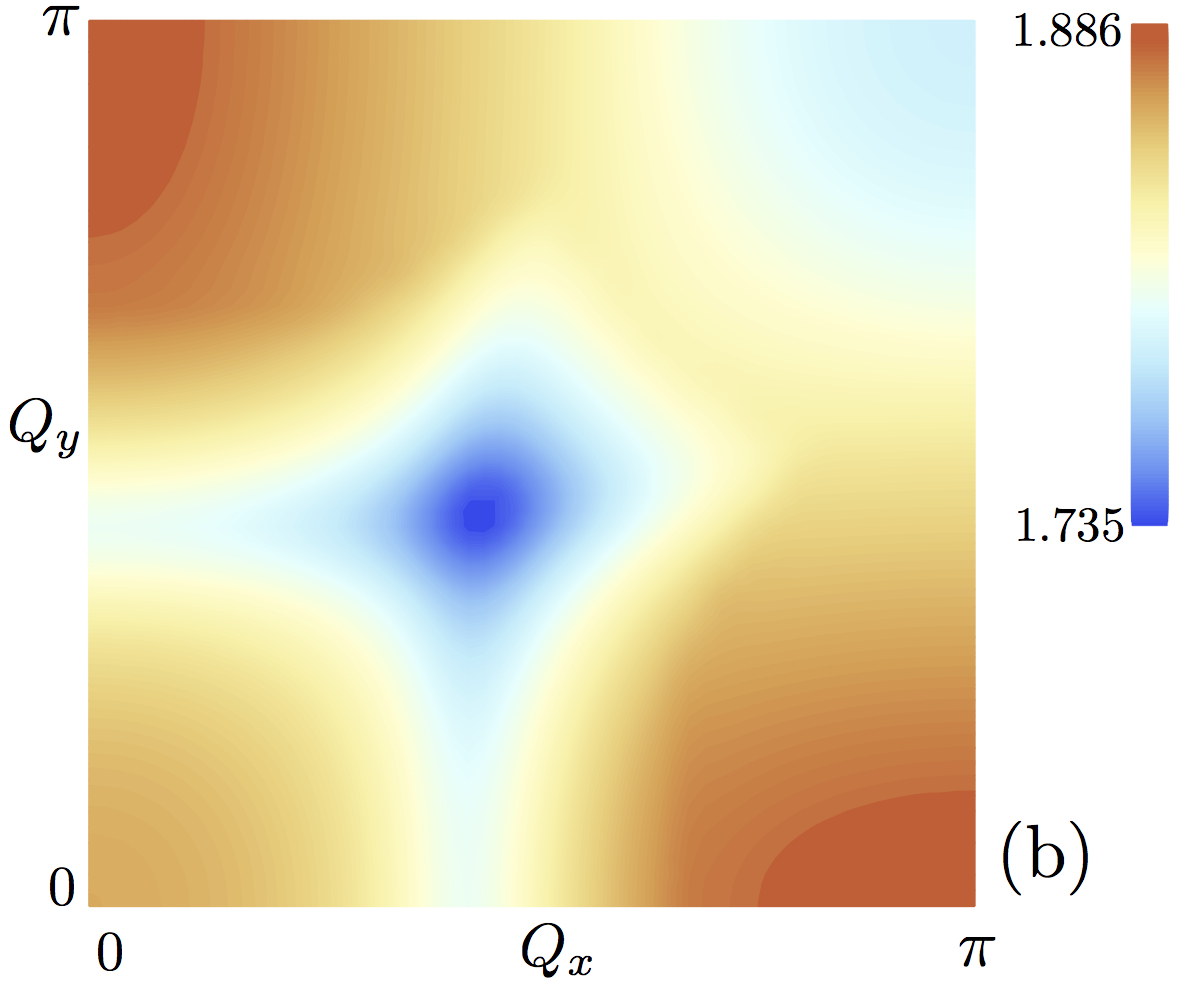}
\includegraphics[width=3in]{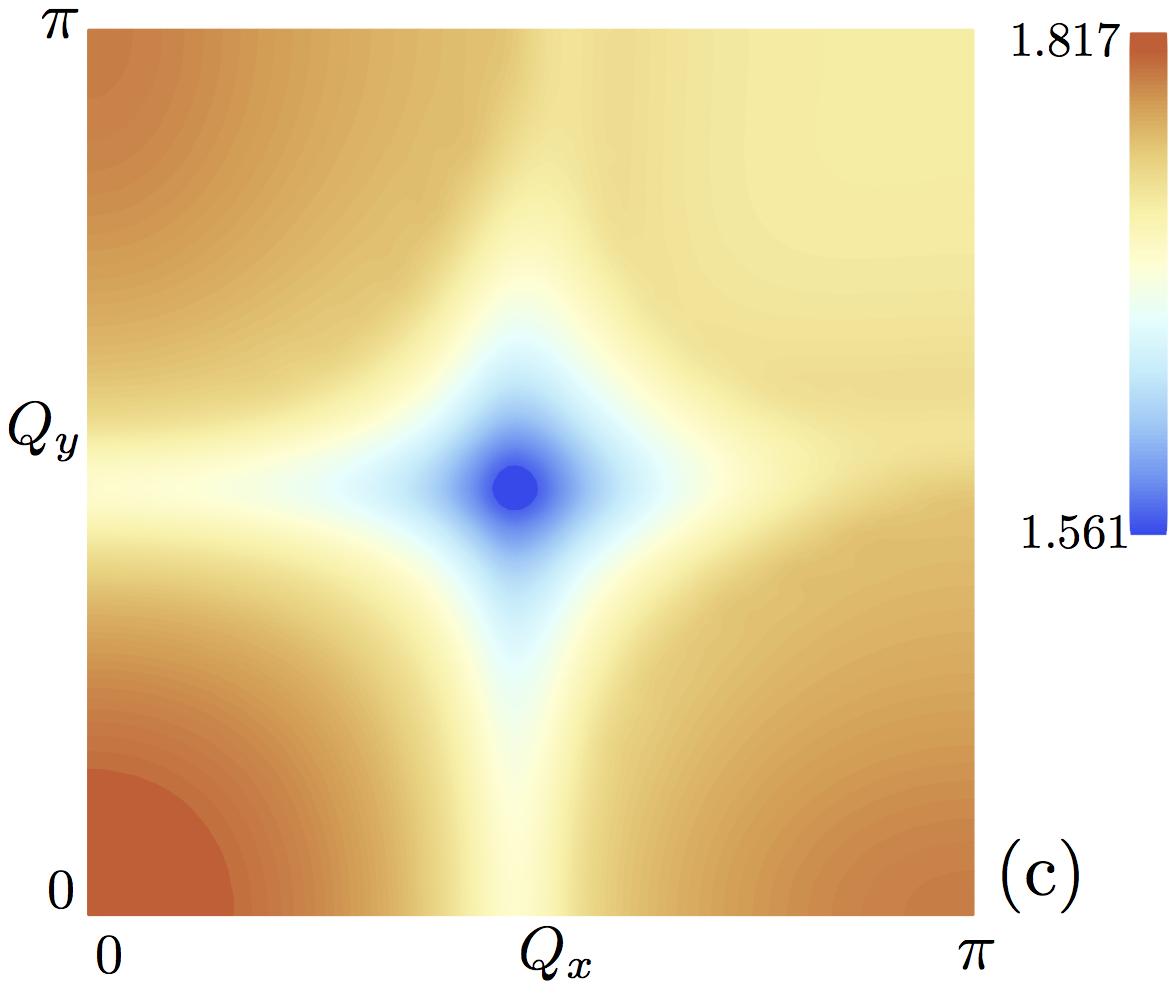}
\caption{As in Fig.~\ref{fig:largensb}, with the $U=\infty$ limit taken via the auxiliary-boson method.
The parameters for the three figures are the same as those in Ref.~\onlinecite{paper1}:
we have $t_1
  = 1$, $t_2\in\{0.5,\,0.16,\,0.18\}$, $t_3\in\{0.6,\,0.9,\,1.6\}$, $J_1
  \in\{0.09,\, 0.235,\,0.4\}$, $V_1 \in\{1.,\,1.5,\,0.5\}$, $J_2=J_3=V_2=V_3=0$.
  The hole density in all three figures is 0.1, corresponding to $\mu \in \{-0.5256,\, -0.90285,\,-1.1174\}$.
}
\label{fig:gutz}
\end{figure}

For completeness, we present in Fig.~\ref{fig:gutz}
the auxiliary-boson results for precisely the same parameters used in Ref.~\onlinecite{paper1}
for the Gutzwiller projected variational wavefunctions. 
In moving from (a) to (c), we find increasing preference for the $(Q_0, Q_0)$ instability, as in Ref.~\onlinecite{paper1}.
However, in (a) the global eigenvalue minimum is for the staggered flux state at $(\pi, \pi)$, while in Ref.~\onlinecite{paper1}
is was for the experimentally observed $(Q_0,0)$ state. Ref.~\onlinecite{paper1} had the staggered flux state preferred in (b),
while here we find that charge order at $(Q_0,Q_0)$ is preferred. 

\section{DMFT approach for large $U$}
\label{sec:dmft}
In this section we present results of an alternative approach to describe the
strong local repulsion. We first perform a dynamical mean field (DMFT)
calculation \cite{GKKR96}
for the tight-binding model with dispersion $\varepsilon_{\bk}$ for a certain
filling factor and value of the interaction $U$. We use the resulting
$\bk$-independent self-energy $\Sigma(i\omega_n)$ to compute the instability
matrix [cf. Eq.~({\ref{defPilm})] related to the $J$-interaction, 
\begin{equation}
  \Pi_{{\rm DMFT},m,n}(\bQ)= \sum_{\bk}\phi_n(\bk)\Pi_{\rm
    DMFT}(\bk,\bQ)\phi_m(\bk),
\end{equation}
where
\begin{eqnarray*}
  \Pi_{\rm
    DMFT}(\bk,\bQ)&=&-\frac{1}{\beta}\sum_nG(i\omega_n,\bk+\bQ/2)G(i\omega_n,\bk-\bQ/2)
  \\
&=&-\frac{1}{\beta}\sum_n
\frac{1}{i\omega_n-\varepsilon_{\bk+\bQ/2}+\mu-\Sigma(i\omega_n)}\frac{1}{i\omega_n-\varepsilon_{\bk-\bQ/2}+\mu-\Sigma(i\omega_n)}.
\end{eqnarray*}
This can be used to anlyze the instability in an equation analogous to Eq.~(\ref{teqn}).
Such a calculation leads a renormalization of the low energy dispersion by a
renormalization factor $z=[1-{\rm Re}\Sigma'(0)]^{-1}$, $\tilde t_i=z t^0_i$, similar to the auxiliary-boson
calculation. It additionally accounts for damping effects of the excitations
away from the Fermi surface and  a split into low energy dispersion and
Hubbard bands. In order to project out double occupancy
completely, one should perform the DMFT calculation at $U\to\infty$. However,
this leads to very small renormalization factors $z$, \cite{ZHPMGS13} at odds
with experimental observations.\cite{Gea07,MJMS07} 
We therefore prefer to perform the calculation for values of $U\sim 1-1.5 W$,
where $W$ is the bandwidth of the tight-binding model. Double occupancy is
reduced to less than 0.05 in such calculations. There is no problem of double counting in
this procedure since the $J$-interaction is absent in paramagnetic DMFT
calculations.\cite{GKKR96}  The DMFT self-consistency problem is solved with the numerical
renormalization group\cite{BCP08} at low temperature. 
The result of such a calculation for $J_1=0.5 $ and filling factor $n=0.85$
are displayed in Fig.~\ref{fig:evn0.85dmft}. 

\begin{figure}
\includegraphics[width=5in]{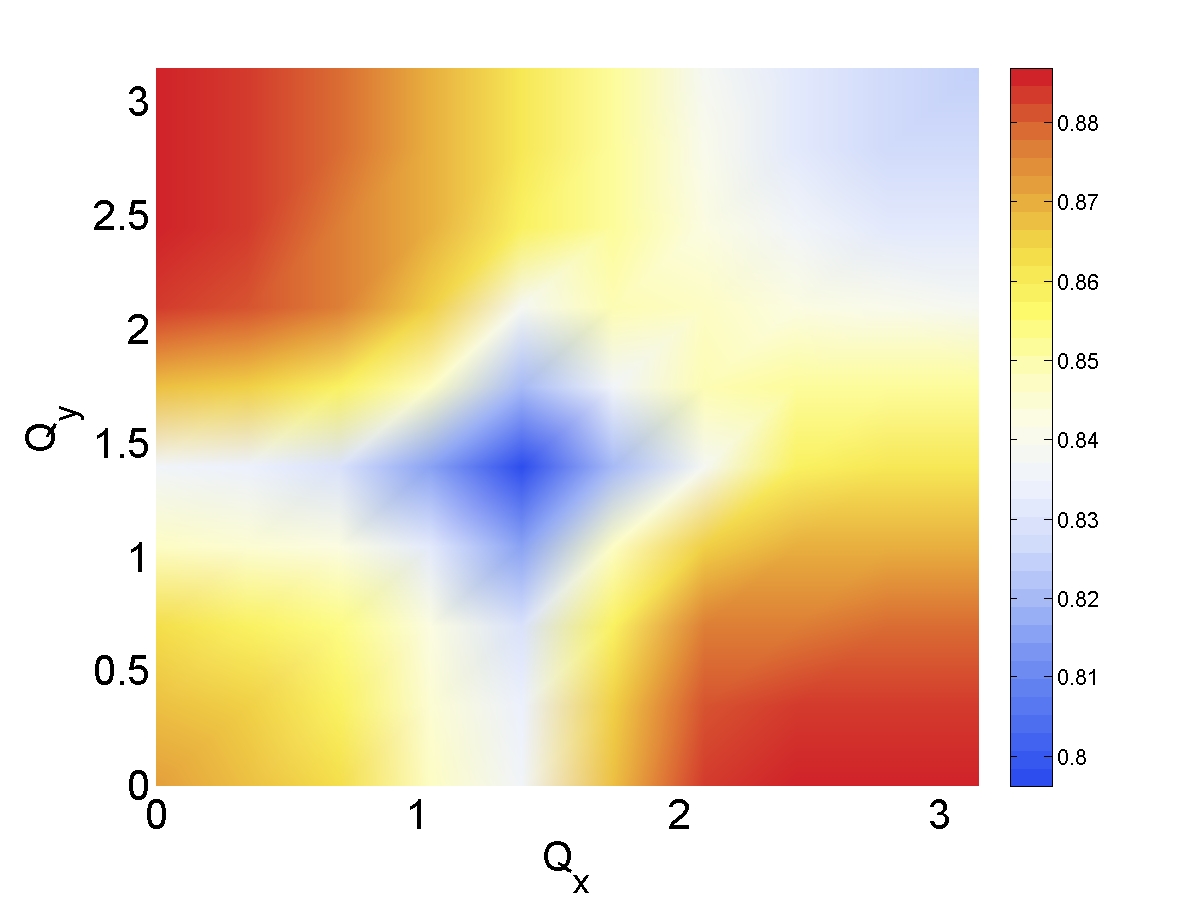}
\caption{Plot of the lowest eigenvalues $\lambda_{\bQ}$ of the instability equation for different
  $\bQ$. The parameters are $T=0.01$, $n=0.85$, $U=8 t^0_1$, $V_i=0$ and the hopping
  parameters as before in Fig.~\ref{fig:bz}. The structure of the eigenvalues and eigenvectors is
  very similar to Fig.~\ref{fig:largensb}.} 
\label{fig:evn0.85dmft}
\end{figure}
As before the dominant instability is at $(Q_0 ,Q_0)$ with subdominant
instabilities at $(Q_0 ,0)$ and $(\pi, \pi)$, and the eigenfunctions are as
discussed above. The value of $Q_0\simeq 0.44\pi$ is
a bit larger than what is expected from the Fermi surface geometry (see
Fig.~\ref{fig:bz}), where for the parameters $Q_0\simeq 0.39\pi$. We have
restricted the analysis here to only finite $J_1$ such that the relevant basis
functions are $\phi_n(\bk)$ with $n=1,2,7,8$. Note that the strength of the
instability is reduced by the renormalization factor $z\simeq 0.25$ which also
acts like a quasipartice weight. For other filling factors and interactions
$U\sim 1.5W$ we find similar results as in Fig.~\ref{fig:evn0.85dmft}. It is
worth noting that at higher temperatures the global minimum can shift to
$(\pi,\pi)$.
We conclude that the structure of dominant charge/bond ordering instabilities
obtained from treating Mott correlations with DMFT is very similar to the
results in Section~\ref{sec:sb}.

\section{Conclusions}

Our main conclusion is that Mott correlations, as implied by the auxiliary-boson and DMFT methods, 
do not significantly modify the conclusions of Ref.~\onlinecite{rolando}.
As long as the metallic state has ``hot spots'' on its Fermi surface, its dominant instability in the spin-singlet,
particle-hole channel is towards a bond-ordered state near wavevectors $(\pm Q_0, \pm Q_0)$ with a local $d$-wave symmetry
of bond ordering; such a state has also been called an ``incommensurate nematic''.
However, our present computations do show an enhanced  instability towards a time-reversal symmetry breaking state with spontaneous
currents: the ``staggered flux'' state.

The experimentally observed charge ordering at $(\pm Q_0,0)$ and $(0, \pm Q_0)$ remained subdominant
to ordering at $(\pm Q_0, \pm Q_0)$. Nevertheless, our computations do predict a predominantly $d$-wave form
for the order parameter $P_\bQ (\bk)$ at $\bQ = (\pm Q_0,0)$ and $(0, \pm Q_0)$, 
as shown in Eqs.~(\ref{splusd}) and (\ref{splusd2}).
We note the variational computations in Ref.~\onlinecite{paper1}, using a 
wavefunction with double occupancy projected out, did find a regime in which the dominant charge ordering was
at $(\pm Q_0, 0)$ and $(0, \pm Q_0)$. Other mechanisms for selecting the observed wavevector
have also been proposed.\cite{efetov,debanjan} 

Finally, we mention two recent experimental reports \cite{comin2,dcdw} concluding that the 
charge order at $(Q_0,0)$ is predominantly $d$-wave, {\em i.e.\/} the $\ell=1$
coefficient of the basis functions $\phi_{\ell} (\bk)$ in Table~\ref{tab:basisfunc} is significantly
larger than all other $\ell$. This is just as in Eqs.~(\ref{splusd}) and (\ref{splusd2}).

\acknowledgments

We thank D. Chowdhury, A. Georges, and J. Sau for valuable discussions.
The research was supported by the U.S.\ National Science Foundation under
grant DMR-1103860, and by the Templeton Foundation. JB acknowledges financial
support from the DFG through grant number BA 4371/1-1.


\begin{thebibliography}{}

\bibitem{paper1} A.~Allais, J.~Bauer, and S. Sachdev, arXiv:1402.4807.

\bibitem{stripes}  S.~A.~Kivelson, I.~P.~Bindloss, E.~Fradkin, V.~Oganesyan, J.~M.~Tranquada, A.~Kapitulnik, and C.~Howald,
Rev. Mod. Phys. {\bf 75}, 1201 (2003).

\bibitem{ssrmp} S.~Sachdev, Rev. Mod. Phys. {\bf 75}, 913 (2003).

\bibitem{vojta4} M.~Vojta and O.~R\"osch, 
Phys. Rev. B {\bf 77}, 094504 (2008).

\bibitem{dhlee} J.~C.~S\'eamus Davis and Dung-Hai Lee, Proc. Natl. Acad. Sci. {\bf 110}, 17623 (2013).

\bibitem{YK00}
H. Yamase and H. Kohno,
J. Phys. Soc. Jpn. {\bf 69}, 2151 (2000).

\bibitem{HM00}
C. J. Halboth and W. Metzner, Phys. Rev. Lett. {\bf 85}, 5162 (2000).

\bibitem{OKF01}
V. Oganesyan, S. A. Kivelson, and E. Fradkin,
Phys. Rev. B {\bf 64}, 195109 (2001).

\bibitem{marston} I. Affleck and J. B. Marston, Phys. Rev. B {\bf 37}, 3774 (1988).

\bibitem{kotliar}  Z.~Wang, G.~Kotliar, and X.-F.~Wang, Phys. Rev. B {\bf 42}, 8690 (1990).

\bibitem{sudip} S.~Chakravarty, R.~B.~Laughlin, D.~K.~Morr, and C.~Nayak, Phys.
Rev. B {\bf 63}, 094503 (2001).

\bibitem{leewen} P. A. Lee, N. Nagaosa, and X.-G. Wen, Rev. Mod. Phys. {\bf 78}, 17 (2006).

\bibitem{laughlin} R.~B.~Laughlin, Phys. Rev. B {\bf 89}, 035134 (2014).

\bibitem{varma} M. E. Simon and C. M. Varma, Phys. Rev. Lett.  {\bf 89}, 247003 (2002).


\bibitem{rolando} S.~Sachdev and R.~La Placa, Phys. Rev. Lett. {\bf 111}, 027202 (2013).

\bibitem{jay}  J. D. Sau and S. Sachdev, Phys. Rev. B {\bf 89},  075129 (2014).

\bibitem{comin} R.~Comin {\em et al.\/} Science {\bf 343} 390 (2013).

\bibitem{grilli} M. Grilli and B. G. Kotliar, Phys. Rev. Lett. {\bf 64}, 1170 (1990)

\bibitem{wang} Z. Wang, G. Kotliar, and X.-F. Wang, Phys. Rev. B {\bf 42}, 8690 (1990).


\bibitem{GKKR96}
A.~Georges, G.~Kotliar, W.~Krauth, and M.~Rozenberg,
Rev. Mod. Phys. {\bf 68}, 13 (1996).

\bibitem{BCP08}
R.~Bulla, T.~Costi, and T.~Pruschke,
Rev. Mod. Phys. {\bf 80}, 395 (2008).


\bibitem{Gea07}
J.~Graf, G.-H. Gweon, K.~McElroy, S.~Y. Zhou, C.~Jozwiak, E.~Rotenberg,
  A.~Bill, T.~Sasagawa, H.~Eisaki, S.~Uchida, H.~Takagi, D.-H. Lee, and
  A.~Lanzara,  Phys. Rev. Lett. {\bf 98}, 067004 (2007).

\bibitem{MJMS07}
A.~Macridin, M.~Jarrell, T.~Maier, and D.~J. Scalapino,
 Phys. Rev. Lett. {\bf 99}, 237001 (2007).

\bibitem{ZHPMGS13}
R.~\ifmmode~\check{Z}\else \v{Z}\fi{}itko, D.~Hansen, E.~Perepelitsky,
  J.~Mravlje, A.~Georges, and B.~S. Shastry,
 Phys. Rev. B {\bf 88}, 235132 (2013).


\bibitem{efetov} H. Meier, C. P\'epin, M. Einenkel, and K. B. Efetov,
arXiv:1312.2010.

\bibitem{debanjan} D.~Chowdhury and S.~Sachdev, to appear.

\bibitem{comin2} R.~Comin {\em et al.\/}, arXiv:1402.5415.

\bibitem{dcdw} K.~Fujita {\em et al.\/}, arXiv:1404.0362.

\end{thebibliography}
\end{document}